\documentclass{iopart}

\usepackage{graphicx}
\usepackage{amsfonts}
\usepackage{color}
\usepackage{verbatim}

\usepackage{hyperref}

\newcommand{\doi}[2]{\href{http://dx.doi.org/#1}{#2}}
\newcommand{\affiliation}[1]{\address{#1}}
\newcommand{\ii}{{\rm i}}
\begin{document}

\title{Inductive dressed ring traps for ultracold atoms}

\author{Matthieu Vangeleyn,$^1$ Barry M.\ Garraway,$^2$ H\'el\`ene Perrin$^3$ and Aidan S.\ Arnold$^1$}
\affiliation{$^1$Department of Physics, SUPA, University of Strathclyde, Glasgow G4 0NG, United Kingdom}
\affiliation{$^2$Department of Physics and Astronomy, University of Sussex, Brighton BN1 9QH, United Kingdom}
\affiliation{$^3$
Laboratoire de physique des lasers, CNRS UMR 7538, Universit\'e Paris 13, Sorbonne Paris Cit\'e, 99 avenue J.-B.\ Cl\'ement, F-93430 Villetaneuse, France}

\date{\today}

\begin{abstract}

We present two novel dressed inductive ring trap geometries, ideal for atom interferometry or studies of superfluidity and well-suited to utilisation in atom chip architectures. The design permits ring radii currently only accessible via near-diffraction-limited optical traps, whilst retaining the ultra-smooth magnetic potential afforded by inductive traps. One geometry offers simple parallel implementation of multiple rings, whereas the other geometry permits axial beam-splitting of the torus suitable for whole-ring atom interferometry.

\end{abstract}

\maketitle

\section{Introduction}

A toroid is one of the simplest multiply-connected 3D shapes, and the periodic boundary conditions both simplify and enrich experimental and theoretical work. In the limit of small cross-sectional area, annular traps permit approximate access to ideal 1D infinite systems (cf.\ Born-von Karman boundary conditions) using a finite experimental footprint. Ring traps are excellent systems in which to perform atom interferometry \cite{zawadzki,gardiner,hadzi} or study superfluidity \cite{hadzi2,phillips3}. Extremely smooth magnetic ring traps for ultracold atoms can be made via induction \cite{Arnold:2008:a,Riis2012}, which eliminates both wire end-effects and roughness from meandering DC currents \cite{Bouchoule:2007:a}. Such time-averaged traps are limited to radii $\geq 5\,$mm for the `thin-wire' geometries considered here, however careful optimisation of wire thickness can decrease radii to $\approx 1\,$mm \cite{griff}. In this paper we show -- for the first time -- how inductive ring traps can be obtained using RF-dressing, which then permits access to the $\leq 0.3\,$mm ring radius regime, suitable for fabrication on atom chips. Such chips offer the prospect of portable ultracold atomic setups and there have been recent important developments in both chip loading \cite{chipload} and their application \cite{chipapplication,chipapplication2}. We envisage prospects for low-decoherence on-chip studies of Sagnac interferometry, superfluidity, ring dark solitons \cite{toikka} as well as vortices \cite{vortex} and solitons \cite{soliton} in low-dimensional systems.

There have been a wide variety of studies of cold matter in toroidal geometries, which we briefly summarize, considering first complementary traps, before concentrating on the purely magnetic techniques relevant to this article.

The optical dipole force can be used in many ways to confine atom in ring traps: by optically plugging a magnetic trap \cite{Raman:2005:a,phillips}, using static `hollow' Laguerre-Gauss beams \cite{hadzi,hadzi2,phillips3,Dholakia:2000:a,Hennequin:2006:a,Fatemi:2007:a,phillips2} or beams that have intensity profiles spatially-shaped by a spatial light modulator \cite{cassett}. In addition, one can quickly scan a focused beam \cite{RubinszteinDunlop:2008:a,Arnold:2008:b,Boshier:2009:a} and trap atoms in the time-averaged dipole potential, opening access to a variety of complex geometries including ring lattices. Static dipole traps using both co- \cite{ferris1} and counter-propagating \cite{ferris2} Laguerre Gauss mode superpositions also offer the flexibility of extending to more exotic ring geometries. 

Magnetic ring-shaped traps have been proposed and realised using either static \cite{Chapman:2001:a,Prentiss:2004:a,Riis:2006:a} or time-averaged \cite{Arnold:2004:a,StamperKurn:2005:a,Ifan} magnetic fields.  In purely magnetic traps, to reduce the size of the system, the current-carrying wires -- needed to connect the system to an external electric source -- get closer to the trap region, perturbing the rotational symmetry of the potential. Furthermore, fragmentation of atomic clouds has been seen for magnetically trapped gases lying close to current carrying wires in atom chips \cite{Fortagh:2002:a,Pritchard:2002:a,Pritchard:2003:a,Hinds:2003:b,schummcorr}. This was attributed to corrugations or irregular domain structure in the conductor that deflect the current from the desired path. A way to circumvent this is the use of alternating current to `time-average' away the defects \cite{Bouchoule:2007:a}.

The need to obtain an ultra-smooth ring trap, with no end effects due to input/output wires and the inherent magnetic smoothness of an ac current, led to the proposal of a time-averaged toroidal trap based on a conducting ring driven by magnetic induction \cite{Arnold:2008:a}. This trap has now been experimentally realised \cite{Riis2012}, however although the trap works well at ring radii of a few mm, it is not scalable down into the sub-mm regime required for atom chips. This is because time-orbiting potential (TOP) traps \cite{Arnold:2004:a,petrich} require $\omega_{\rm T} \ll \omega_{\rm TOP} \ll \omega_{\rm L}$ where the subscripts T, TOP and L refer to the harmonic trap, rotating bias field, and Larmor angular frequencies at the trap location, respectively. Ring-shaped conductors with radius $r_r$ and wire radius $r_w$ have a natural high angular frequency cutoff $\omega_{RL} = R/L \propto (\ln(8 r_r/r_w)-1.75)\,{r_w}^{-2}$, however we have shown that inductive TOP traps work best with $\omega_{\rm TOP}$ a few times greater than $\omega_{RL}$ \cite{Arnold:2008:a}. For a given wire geometry $(r_w/r_r)$ this leads to a prohibitive increase in $\omega_{\rm TOP}$ (above the Larmor angular frequency) due to the rapid increase in $\omega_{RL}$ as the ring size shrinks. Throughout this paper we consider $r_w/r_r=0.07$, suitable for chip traps, however  $\omega_{RL}$ can be reduced to some extent by increasing $r_w$ for a given $r_r.$

Recently, using radio-frequency adiabatic potentials \cite{Garraway:2001:a}, new ways to create toroidal ring traps have been suggested
\cite{Garraway:2006:a,Schmiedmayer:2006:a,Spreeuw:2007:a,vonKlitzing:2007:a,Foot:2008:a}  offering extended flexibility for trap geometries. In particular a time-averaged adiabatic potential (TAAP) \cite{vonKlitzing:2007:a} ring trap has recently been experimentally demonstrated \cite{Foot:2011}. In this paper, we present a new scheme for creating a toroidal trap with a radius in the 100~$\mu$m range with magnetic fields oscillating at RF frequencies, extending the inductive trap from the TOP regime \cite{Arnold:2008:a}, into the higher-frequency adiabatic regime, whilst retaining all the benefits of the inductive geometry. Magnetic traps are inherently free of the wavelength-scale diffraction corrugation that can potentially plague optical traps with radii $\geq 20\,\mu$m \cite{hadzi2,phillips3}, and as our trap design is suitable for implementation on atom chips, it will have advantages over dressed traps with macroscopic coils \cite{Foot:2011} in terms of gradiometry and portability.

\section{General theory of dressed inductive ring traps}

We first present the central principles of the dressed inductive ring trap, then discuss two different implementable layouts, i.e.\ with the quantisation magnetic field being either spatially uniform but with time-varying direction (a TAAP-type trap) or a static quadrupole configuration. To begin generally, we consider an atom in a static or slowly varying quantisation magnetic field ${\bf B}_{\rm S}({\bf r},t)$ with an alternating magnetic field ${\bf B_{\rm RF}}({\bf r})$ at angular frequency $\omega$. In the dressed state picture, the AC field induces a coupling of the atomic Zeeman sub-levels and the resulting interaction Hamiltonian can be written
\begin{equation}
    H({\bf r},t)=\Omega_{\rm RF}({\bf r}) \rm{F_{\sigma^s}}\cos{(\omega t)}+\Omega_{\rm S}({\bf r},t)\rm{F_Z},
\end{equation}
where $\rm{F_{\sigma^s}}$ and $\rm{F_Z}$ are respectively the projections of the total angular momentum operator along the $\sigma^s$ polarisation and $Z$ directions (if we choose a local co-ordinate system with $Z$ parallel to ${\bf B}_{\rm S}$). The parameter $s={\rm sign}(g_F)$ is the sign of the Land\'{e} g-factor $g_F$ for the total angular momentum $F$. The Larmor frequency is expressed as $\Omega_{\rm S}({\bf r},t)=g_F \mu_B B_{\rm S}({\bf r},t) / \hbar$ and $\Omega_{\rm RF}({\bf r})=g_F \mu_B B_{\sigma^s}({\bf r})/\hbar$ denotes the Rabi frequency associated with the part of the AC field that contributes to the coupling between the Zeeman sub-levels, where $\mu_B$ is the Bohr magneton. In this limit, one can simply express the resulting adiabatic potential as
\begin{equation}
    U({\bf r},t)= \hbar \, m_F \sqrt{\delta^2({\bf r},t)+{\Omega_{\rm RF}}^2({\bf r})}=\hbar \, m_F \Omega,
    \label{eq:AdiabPot}
\end{equation}
for an atom initially in the $m_F$ Zeeman sub-level, with $\delta({\bf r},t)=\Omega_{\rm S}({\bf r},t)-\omega$ the detuning between the Larmor and drive frequencies. These expressions for $H({\bf r},t)$ and $U({\bf r},t)$ stand in the adiabatic regime, i.e.\ as long as the condition $|\theta'({\bf r},t)|\ll \Omega({\bf r},t)$ is valid, where $\theta=\arctan(\Omega_{\rm RF}/\delta)$. Adiabaticity can thus be re-expressed as
 \begin{equation}
   A =  \frac{\left(\delta^2+{\Omega_{\rm RF}}^2\right)^{3/2}}{ \left|\delta \, {\Omega_{\rm RF}}' - {\delta}' \, \Omega_{\rm RF}\right|} \gg 1,
\label{lz}
 \end{equation}
where $'$ represents the total time derivative $\frac{d}{dt}=\frac{\partial}{\partial t}+{\bf v}\cdot {\bf \nabla},$ allowing for atomic motion at velocity ${\bf v},$ and we refer to $A$ as the adiabaticity parameter. We will only consider trapping regions in which the rotating wave approximation (RWA) is valid, i.e.\ $\eta_\delta=|\delta({\bf r},t)|/(\Omega_{\rm S}({\bf r},t)+\omega) \ll 1,$ and $\eta_{\rm RF} = \Omega_{\rm RF}/\omega \ll 1$ \cite{schmiedRWA}.

\begin{figure}[!b]
\centering\includegraphics[width=\columnwidth]{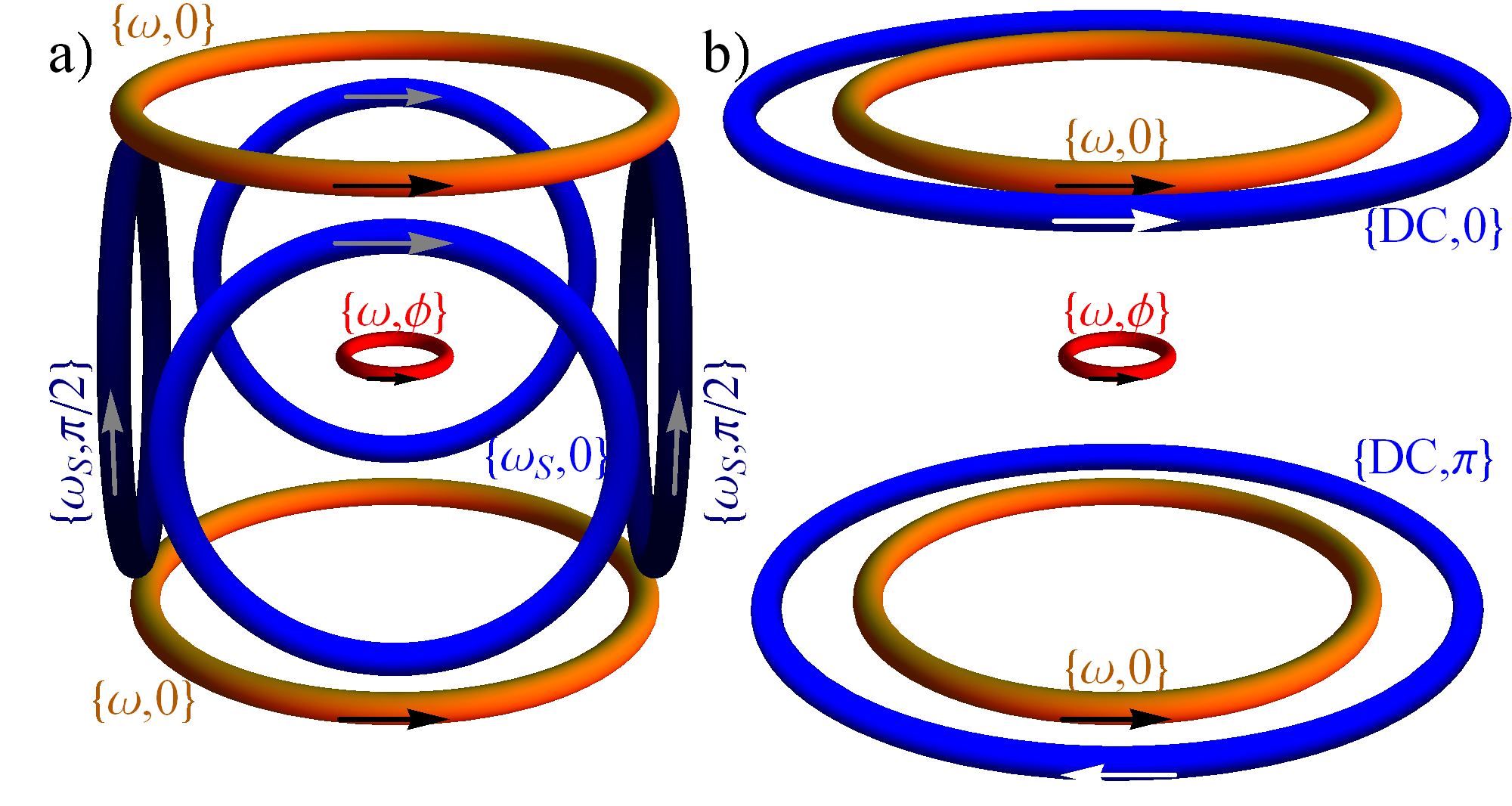}
    \caption{Magnetic coil schematics. In both geometries, the driving ac-Helmholtz coils (orange) and the ring in which current is magnetically induced (red) oscillate at angular frequency $\omega$ and have axes aligned with the $z$ direction. The two geometries we consider are: a) two Helmholtz coil pairs (blue and dark blue) driven in quadrature at an angular frequency $\omega_{\rm S} \ll \omega$, b) a DC anti-Helmholtz coil pair (blue). Black, gray and white arrows indicate current directions at angular frequencies $\omega,$ $\omega_{\rm S}$ and DC, respectively. Co-ordinate pairs are coil current angular frequencies, with associated phases.}
    \label{fig:gRingSchem}
\end{figure}

Consider now the experimental apparatus illustrated in Fig.~\ref{fig:gRingSchem}: a set of two Helmholtz coils (orange) provide a spatially homogenous oscillating field $B_{\rm H}\,\hat{{\bf z}}\,e^{\ii \omega t}$ throughout the surface of the conductive ring (red), and parallel to the ring axis $(z)$. Spatial homogeneity of $B_{\rm H}$ is reasonable if we consider the ring to have much smaller dimensions than the coils, and homogeneity is not essential as long as cylindrical symmetry is not significantly broken. The magnetic flux from $B_{\rm H}$ induces a current in the ring, which has a complex amplitude calculated using Lenz's law and characterised using the ring's electrical resistance $R$ and inductance $L$ \cite{Arnold:2008:a}:
\begin{equation}
    I_{\rm Ring}=-\frac{\pi {r_{r}}^2 B_{\rm H}}{L(1-\ii \,\omega_{RL}/\omega)}=|I_{\rm Ring}| \, e^{\ii \phi},
\end{equation}
where $r_r$ is the radius of the ring. The above formalism using complex currents allows one to drop time dependence.  This current, in turn, generates a synchronous magnetic field ${\bf B}_{\rm ring}({\bf r})  e^{\ii \phi}$ which is fully spatially expressed using elliptical integrals \cite{Paul:book:a}. The complete RF field, with time dependence, at position $\textbf{r}$ is thus:
\[{\bf B}_{\rm RF}({\bf r})=\left({\bf B}_{\rm ring}({\bf r})\; e^{\ii \phi} +B_{\rm H}\,\hat{{\bf z}}\right)\,e^{\ii \omega t}.\] Now we consider a static, or slowly varying, magnetic field ${\bf B}_{\rm S}({\bf r})$, which we choose as our quantisation axis. The RF field has to be expressed in terms of the two components $B_{\perp_1}({\bf r})$ and $B_{\perp_2}$({\bf r}) orthogonal to the static field, since the parallel $(\pi)$ component does not contribute to coupling. Using Jones' formalism \cite{Jones}, the amplitude of the RF fields that couples to $\sigma^{\pm}$ transitions is then given by:
\begin{equation}
    B_{\sigma^{\pm}}({\bf r})=\frac{1}{2\sqrt{2}}\Big|(1\pm \ii)B_{\perp_1}({\bf r})+(1\mp \ii)B_{\perp_2}({\bf r})\Big|.
\end{equation}
It is then straightforward to calculate the adiabatic potential using Eq.~\ref{eq:AdiabPot}.

Throughout this paper, unless otherwise stated, we assume a copper ring with a radius of $400\,\mu$m and a wire radius of $28\,\mu$m. Helmholtz coils provide an RF field of amplitude $B_{\rm H}=12\,$G oscillating at $\omega=2\pi\times10\,$MHz. Our simulations include a finite-element solver \cite{Riis2012} to account for the finite ring size and the skin effect that can significantly change the electrical parameters at high frequencies, as well as for precise calculation of the potential depth. Note, however, that the main results of the paper can still be reproduced to reasonable ($\approx 10\%$) accuracy without resorting to finite-element calculations. With the above ring dimensions and driving field, the ring has resistance and inductance of $18\,$m$\Omega$ and $1.5\,$nH, respectively. The current amplitude flowing inside the conductor is $390\,$mA, dissipating about $1.4\,$mW that rapidly leads to an equilibrium temperature of the copper of approximatively 230$^\circ$C (assuming perfect black body radiation). If required, this temperature rise could be radically reduced using e.g.\ a diamond substrate as an electrically insulating heatsink. We also consider the specific case of $^{87}$Rb pumped into the $|F=2,\,m_F=2\rangle$ magnetic sub-level, therefore $g_F=1/2$ and only the $B_{\sigma^+}({\bf r})$ component couples to the atomic spin.

\section{Geometry 1: Inductive TAAP ring}

\begin{figure}[!t]
\begin{minipage}{.783\columnwidth}
   \includegraphics[width=\columnwidth]{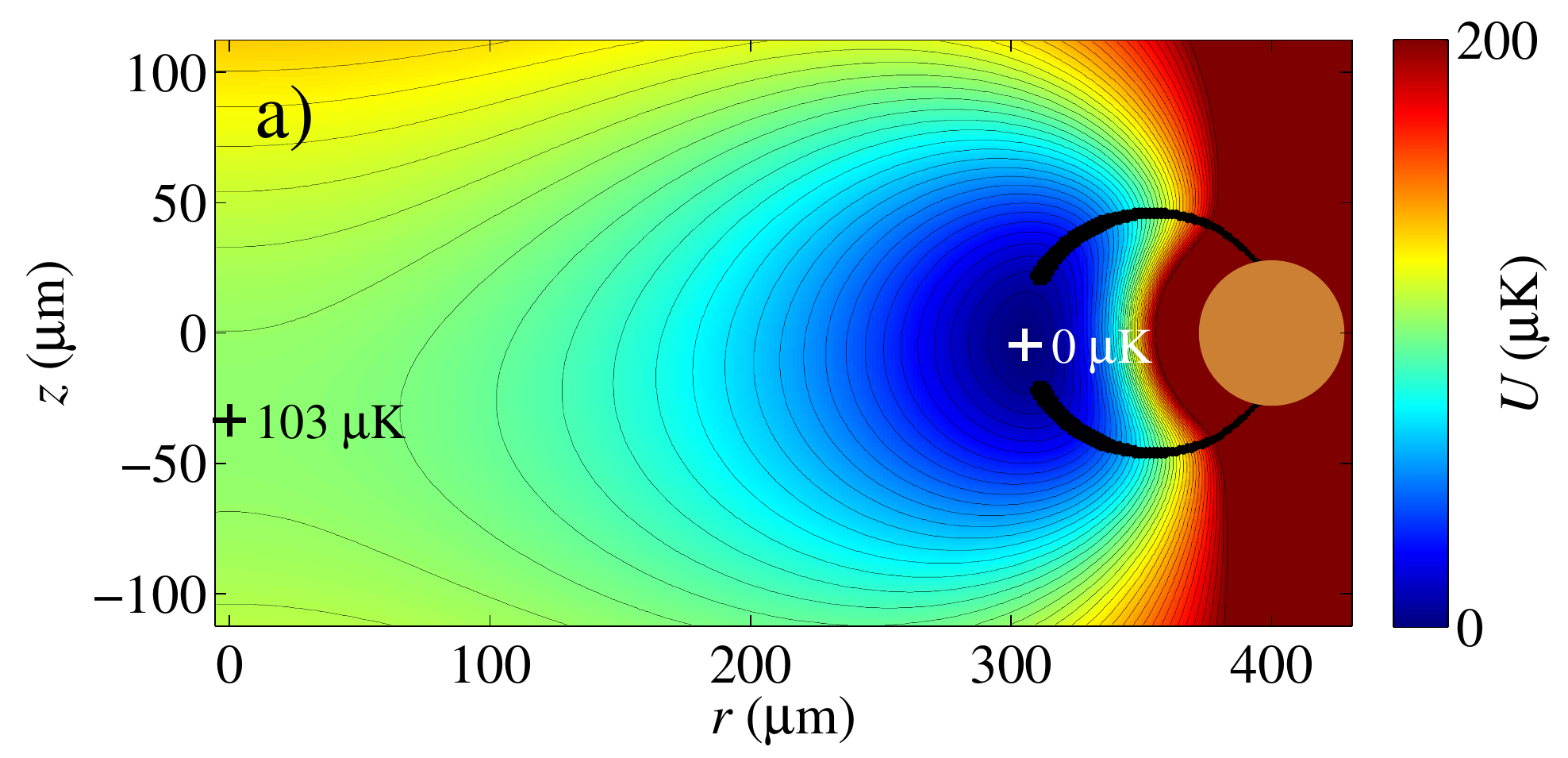}
\end{minipage}
\begin{minipage}{.21\columnwidth}
   \includegraphics[width=\columnwidth]{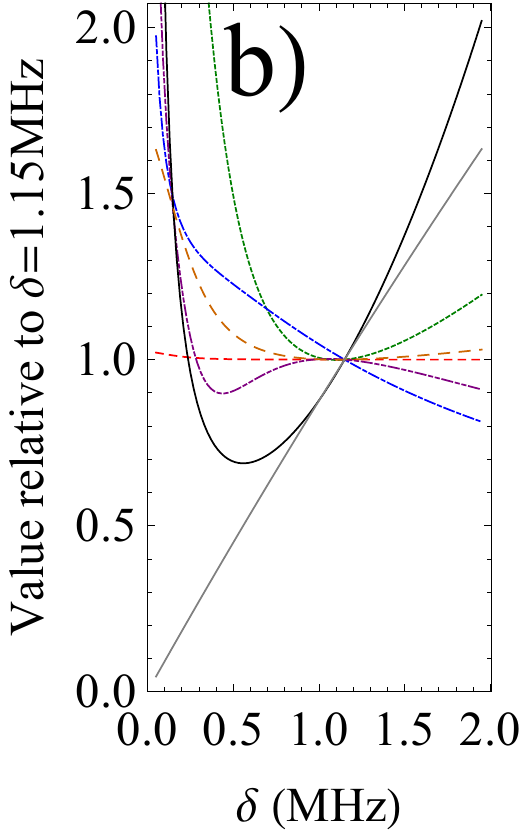}
\end{minipage}
    \caption{
a) The time-averaged potential resulting from rotating the quantisation field in the $xy$ plane of the ring. White and black crosses indicate trap minima and saddle points, respectively, and the ring wire is shown by the copper disk. The black regions show the aggregate of the zones where the RF coupling drops below $2\pi \times 100\,$kHz as the quantisation axis angle varies in the range 0 to $2 \pi$ (see also the \href{http://photonics.phys.strath.ac.uk/wp-content/uploads/2014/02/Fig2Movie.gif}{\textbf{supplementary movie}} of the instantaneous potential as the quantisation field rotates). The colour scale is relative to the trap minimum, black contours are 5~$\mu$K isopotential lines. b) The relative change of trap parameters, compared to the $\delta=1.15\,$MHz case (see values in main text), as the dressing detuning $\delta$ is varied for: radial (blue, dash-dotted) and axial (purple, dash-dot-dotted) trap frequencies, trap radius (red, dashed) and axial position (green, dotted), $A$ (black), $\eta_\delta$ (gray) and $\eta_{\rm RF}$ (orange, long-dashed).}


    \label{fig:gQSSchem}
\end{figure}

We now focus on the inductive TAAP configuration (Fig.~\ref{fig:gRingSchem}~a), where the quantisation field is spatially homogenous, with a vector direction in the $xy$ plane of the ring, but rotating around the symmetry axis. For a snapshot in time the cylindrical symmetry of the system is broken by the quantisation field ${\bf B}_{\rm S}$ -- as the total RF field $B_{\rm RF}({\bf r})$ is projected into a different ratio of coupling:non-coupling components, depending on the quantisation direction. However, as the atoms can't respond on these time-scales, they experience the cylindrically symmetric time-averaged TAAP potential averaged over one rotation period of the quantisation field shown in Fig.~\ref{fig:gQSSchem}. Here the constant magnetic field amplitude of ${\bf B}_{\rm S}(t)=15.9\,$G creates a corresponding detuning $\delta$ which is spatially and temporally constant, i.e.\ $2\pi\times1.15\,$MHz. The time averaged trap is located 
slightly below the plane of the ring wire due to gravitational sag, with radial and axial trap frequencies of $400\,$Hz and $250\,$Hz, respectively.

There are two factors to consider for the maximum trappable temperature in the inductive TAAP. Firstly, gravity results in a saddle point shown in Fig.~\ref{fig:gQSSchem}, and atoms hotter than $103\,\mu$K will no longer be confined to a ring but have access to a disk. Secondly, one can show that, during the rotation of the quantisation field, there are curves of points where the RF coupling totally vanishes (black zones). However, as the detuning is non-zero when the coupling vanishes, the adiabatic potential still exists. One can estimate the loss at these locations using the $A$ parameter, Eq.~\ref{lz}. These points would in principle limit the trap depth to 5~$\mu$K, however the smallest value of the loss parameter is $A=310$ near the instantaneous 3D point of zero-coupling  (compared to $A=1600$ at the trap minimum) for $100\,\mu$K atoms with a quantisation axis rotation rate of $10\,$kHz. We can therefore expect negligible Landau-Zener loss, so atom temperature and the background pressure will be the only limits to the ring geometry and lifetime, respectively. The RWA is valid at the trap location as $\eta_\delta =0.05$ and $\eta_{\rm RF}=0.07$ (cf.\ experimental parameters in Fig.~2 of Ref.~\cite{Foot:2008:a} where $\eta_\delta =0.12$ and $\eta_{\rm RF}=0.14$).

The trap parameters of the inductive TAAP are amenable to scaling arguments and we consider a scenario where the ring radius and wire radius both scale together with the parameter $\zeta$ (i.e. $r_r \rightarrow \zeta r_r$ and $r_w \rightarrow \zeta r_w$). The natural angular frequency of the ring then transforms as $\omega_{RL} \rightarrow \zeta^{-2} \omega_{RL}$, so if the RF drive angular frequency follows an equal scaling $\omega \rightarrow \zeta^{-2} \omega$, the skin depth of the RF current in the ring scales with $1/\sqrt{\omega}$ i.e. $\zeta$, and the spatial current distribution in the wire remains the same. Moreover, if the amplitude of the RF field $\Omega_{\rm RF}$ and the detuning $\delta$ are kept constant, the TAAP trap potential depth and shape remain the same (if gravitational potential is relatively weak), but are scaled to cover a region $\zeta$ times the original size -- i.e. the radial and axial trap frequencies are $1/\zeta$ times larger.

\section{Geometry 2: Inductive quadrupole ring}

\begin{figure}[!t]
\centering\includegraphics[width=.8\columnwidth]{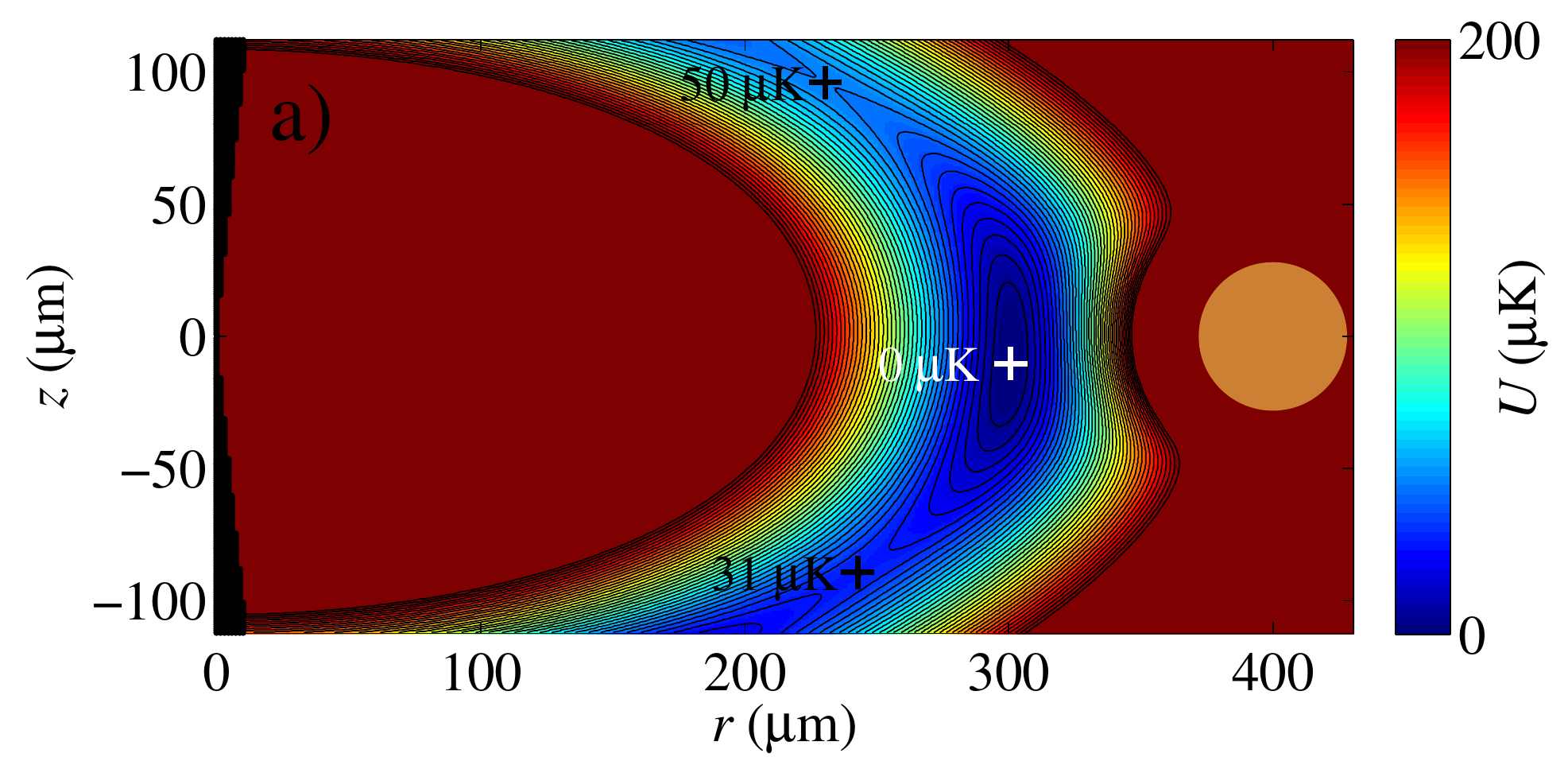}
\centering\includegraphics[width=.8\columnwidth]{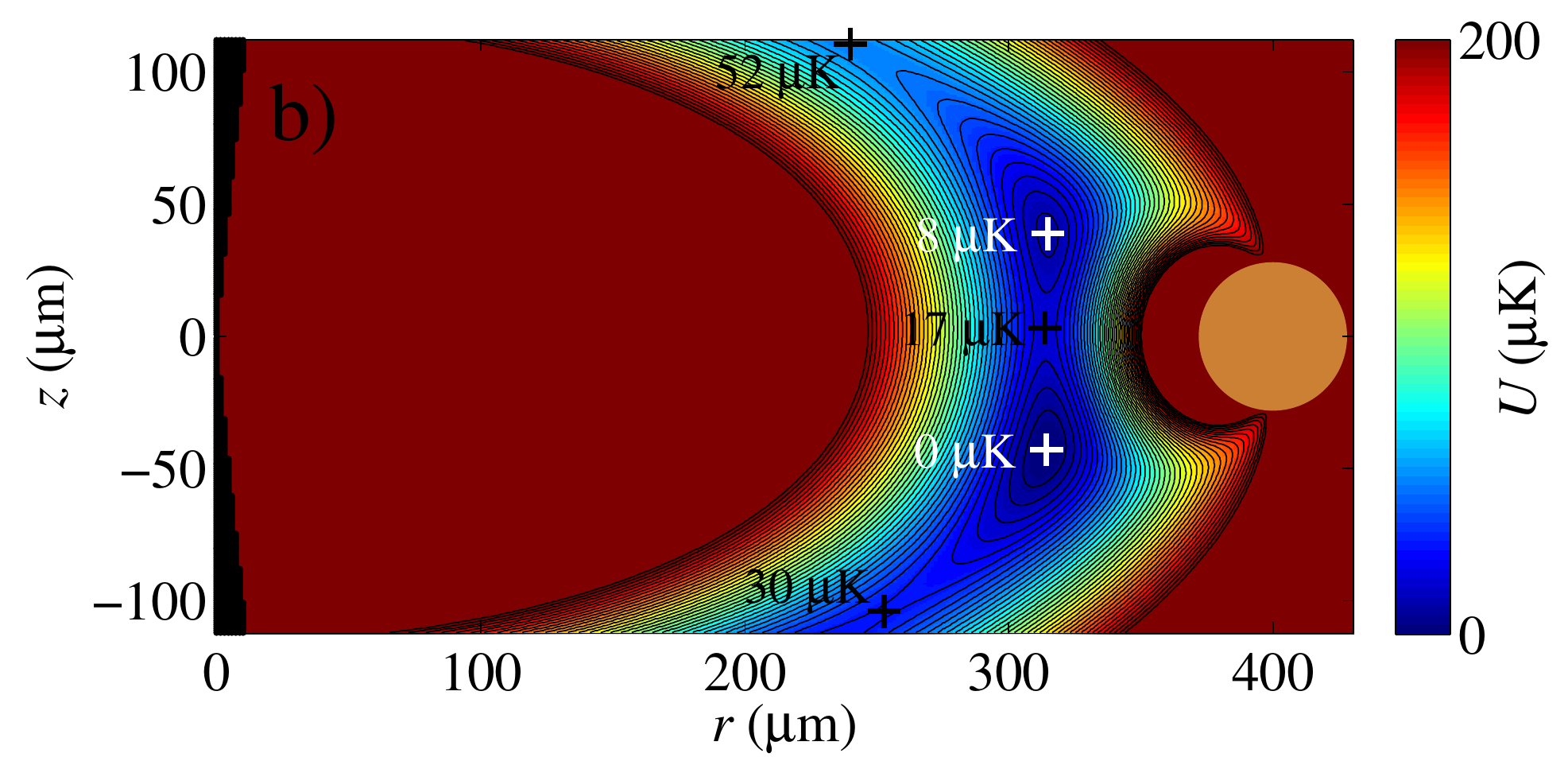}
    \caption{Potentials obtained using a quadrupole quantisation field with a radial gradient of a) $480\,$G/cm and b) $440\,$G/cm, illustrating the transition from a single to a double ring trap. White and black crosses indicate trap minima and saddle points, respectively, and the ring wire is shown by the copper disk. The colour scale is relative to the trap minimum, black contours are 5~$\mu$K isopotential lines. The black region indicates where RF coupling drops below $2\pi \times 100\,$kHz. A \href{http://photonics.phys.strath.ac.uk/wp-content/uploads/2014/02/Fig3Movie.gif}{\textbf{supplementary movie}} shows the axial splitting of the ring trap as the quadrupole gradient is lowered).}
    \label{fig:gQuadSchem}
\end{figure}

We turn now to the second inductive dressed ring configuration (Fig.~\ref{fig:gRingSchem}~b), using a static quadrupole quantisation field, yielding the cylindrically symmetric potentials depicted in Fig.~\ref{fig:gQuadSchem}. Because of the spatially varying magnetic field amplitude, the detuning $\delta$ is spatially dependent. The geometry of the potential is determined by the strength of the magnetic quadrupole. We maintain all other parameters, including $\omega=2\pi\times10\,$MHz, $B_{\rm H}=12\,$G. In the case of a radial quadrupole gradient of $480\,$G/cm, the trap centre is located at $\{r,z\}=\{300,-10\}\,\mu$m, with radial and axial trap frequencies of $750\,$Hz and $180\,$Hz, respectively (Fig.~\ref{fig:gQuadSchem} a). Gravity limits the potential to $\sim 31\,\mu$K. 
Note the only place where the RF coupling vanishes is along the ring axis, far from the trap minimum. The $A$ parameter is $A=1100$ near the trap minimum so we can again expect negligible loss. The RWA is valid at the trap location as $\eta_\delta =0.007$ and $\eta_{\rm RF}=0.07$.

A key advantage of this geometry is that if we change the strength of the quantisation quadrupole to a radial gradient of $440\,$G/cm, one can split the ring into two (Fig.~\ref{fig:gQuadSchem} b), which could be useful for e.g.\ atom interferometrical determination of gravity \cite{chipapplication2}. The simple scaling arguments for the inductive TAAP do not hold for the quadrupole trap, due to the different spatial scaling of $\delta$ and $\Omega_{\rm RF}$. However, with appropriate modifications to the trap parameters, scaling to larger and smaller ring radii is possible.

In terms of loading both kinds of dressed inductive ring trap (TAAP and quadrupole), the relatively shallow depth due to its small dimensions necessitates pre-evaporation of the atomic cloud, after optical molasses, to tens of microKelvin in a magnetic or spin-polarised optical dipole trap. By applying bias fields to a quadrupole/TOP/Ioffe magnetic trap, or focal point scanning of a single-beam dipole trap with an AOM/moving lens, the pre-evaporated cloud could be moved to a localised region of the dressed ring trap location prior to switching it on.

We note that a third dressed inductive geometry is possible, using a DC current-carrying wire along the ring axis (cf.~Ref.~\cite{Riis:2006:a}). This also has very favourable trapping parameters, however we have omitted detailed description due to the difficulty of experimentally realising such a geometry on a chip.

\section{Conclusions}

In conclusion we have extended inductive ring traps from the time-averaged case to the dressed domain, suitable for investigations of sub-millimetre ring traps. Two geometries were considered -- an inductive TAAP ring trap, and a splittable inductive quadrupole ring trap -- both of which are amenable to implementation on atom chips. Inductive TAAPs have the advantage that they are ideally made with spatially homogeneous inducing and quantisation fields, and as such many rings can be implemented in parallel on the same chip for application in e.g.\ gradiometry. Inductive quadrupole traps have quasistatic fields which must be centred on the ring, however multiple quadrupoles for each ring can be implemented on the same chip using U-wires, making parallel implementation of this geometry feasible as well. The inductive quadrupole trap also enables tunable splitting of the ring into a double ring, permitting measurement of e.g. gravitational acceleration, in addition to the usual use of ring traps as gyroscopes. As both types of inductive ring trap have no reliance on optical dipole forces, their potential should be extremely smooth, even with large (mm) scale diameters, ideal for both interferometry and studies of superfluidity.

\section*{Acknowledgements}

Many thanks to Jonathan Pritchard for his finite-element code for the calculation of the ring current distribution. A.S.A.\ and M.V.\ gratefully acknowledge valuable discussions with Erling Riis and Paul Griffin, as well as funding via the Leverhulme Trust and EPSRC grant EP/G026068/1. H.P.\ acknowledges Institut Francilien de Recherche sur les Atomes Froids (IFRAF) for support. B.M.G.\ thanks EPSRC for support through grant EP/I010394/1.


\section*{References}

\begin{thebibliography}{99}


\bibitem{zawadzki}
Zawadzki M E, Griffin P F, Riis E and Arnold A S 2010 \textit{Phys.\ Rev.\ A} 
\doi{10.1103/PhysRevA.81.043608}{\textbf{81}, 043608}.

\bibitem{gardiner}
Halkyard P L, Jones M P A and Gardiner S A 2010 \textit{Phys.\ Rev.\ A} 
\doi{10.1103/PhysRevA.81.061602}{\textbf{81} 061602(R)}

\bibitem{hadzi}
Moulder S, Beattie S, Smith R P, Tammuz N and Hadzibabic Z 2012 \textit{Phys.\ Rev.\ A} 
\doi{10.1103/PhysRevA.86.013629}{\textbf{86} 013629}

\bibitem{hadzi2}
Beattie S, Moulder S, Fletcher R J and Hadzibabic Z 2013 \textit{Phys.\ Rev.\ Lett.\ }
\doi{10.1103/PhysRevLett.110.025301}{\textbf{110} 025301}

\bibitem{phillips3}
Wright K C, Blakestad R B, Lobb C J, Phillips W D and Campbell G K 2013 \textit{Phys.\ Rev.\ Lett.\ }
\doi{10.1103/PhysRevLett.110.025302}{\textbf{110} 025302}

\bibitem{Arnold:2008:a}
Griffin P F, Riis E and Arnold A S 2008 \textit{Phys.\ Rev.\ A} 
\doi{10.1103/PhysRevA.77.051402}{\textbf{77} 051402}

\bibitem{Riis2012}
Pritchard J D, Dinkelaker A N, Arnold A S, Griffin P F and Riis E 2012 \textit{New J.\ Phys.\ }
\doi{10.1088/1367-2630/14/10/103047}{\textbf{14} 103047}

\bibitem{griff}
P. F. Griffin, Private communication.

\bibitem{Bouchoule:2007:a}
Trebbia J-B, Garrido Alzar C L, Cornelussen R, Westbrook C I and Bouchoule I 2007 \textit{Phys.\ Rev.\ Lett.\ }
\doi{10.1103/PhysRevLett.98.263201}{\textbf{98} 263201}

\bibitem{chipload}
Huet L, Ammar M, Morvan E, Sarazin N, Pocholle J-P, Reichel J, Guerlin C and Schwartz S 2012 
\textit{Appl.\ Phys.\ Lett.\ }
\doi{10.1063/1.3689777}{\textbf{100} 121114}; 
Nshii C C, Vangeleyn M, Cotter J P, Griffin P F, Hinds E A, Ironside C N, See P, Sinclair A G, Riis E 
and Arnold A S 2013 \textit{Nature Nanotech.\ }
\doi{10.1038/NNANO.2013.47}{\textbf{8} 321};
Jian B and van Wijngaarden W A 2013 \textit{J.\ Opt.\ Soc.\ Am.\ B} \doi{10.1364/JOSAB.30.000238}{\textbf{30} 238}

\bibitem{chipapplication}
Riedel M F, B\"{o}hi P, Li Y, H\"{a}nsch T W, Sinatra A and Treutlein P 2010 \textit{Nature} 
\doi{10.1038/nature08988}{\textbf{464} 1170}; 
Maineult W, Deutsch C, Gibble K, Reichel J and Rosenbusch P 2012 \textit{Phys.\ Rev.\ Lett.\ }
\doi{10.1103/PhysRevLett.109.020407}{\textbf{109} 020407}

\bibitem{chipapplication2}
Baumg\"{a}rtner F, Sewell R J, Eriksson S, Llorente-Garcia I, Dingjan J, Cotter J P and Hinds E A 2010 
\textit{Phys.\ Rev.\ Lett.\ }
\doi{10.1103/PhysRevLett.105.243003}{\textbf{105} 243003}

\bibitem{toikka}
Toikka L A and Suominen K-A 2013 \textit{Phys.\ Rev.\ A} \doi{10.1103/PhysRevA.87.043601}{\textbf{87}  043601}

\bibitem{vortex}
Madison K W, Chevy F, Wohlleben W and Dalibard J 2000 \textit{Phys.\ Rev.\ Lett. } 
\doi{10.1103/PhysRevLett.84.806}{\textbf{84} 806}

\bibitem{soliton}
Burger S, Bongs K, Dettmer S, Ertmer W, Sengstock K, Sanpera A, Shlyapnikov G V and Lewenstein M 1999
\textit{Phys.\ Rev.\ Lett.\ } \doi{10.1103/PhysRevLett.83.5198}{\textbf{83} 5198}

\bibitem{Raman:2005:a}
Naik D S and Raman C 2005 \textit{Phys.\ Rev.\ A} \doi{10.1103/PhysRevA.71.033617}{\textbf{71} 033617}

\bibitem{phillips}
Ryu C, Andersen M F, Clad\'e P, Natarajan V, Helmerson K and Phillips W D 2007 \textit{Phys.\ Rev.\ Lett. }
\doi{10.1103/PhysRevLett.99.260401}{\textbf{99} 260401}

\bibitem{Dholakia:2000:a}
Wright E M, Arlt J and Dholakia K 2000 \textit{Phys.\ Rev.\ A} 
 \doi{10.1103/PhysRevA.63.013608}{\textbf{63} 013608}

\bibitem{Hennequin:2006:a}
Courtade E, Houde O, Cl\'ement J-F, Verkerk P and Hennequin D 2006 \textit{Phys.\ Rev.\ A} 
 \doi{10.1103/PhysRevA.74.031403}{\textbf{74} 031403}

\bibitem{Fatemi:2007:a}
Olson S E, Terraciano M L, Bashkansky M and Fatemi F K 2007 \textit{Phys.\ Rev.\ A} 
\doi{10.1103/PhysRevA.76.061404}{\textbf{76} 061404}

\bibitem{phillips2}
Ramanathan A, Wright K C, Muniz S R, Zelan M, Hill W T, Lobb C J, Helmerson K, Phillips W D and 
Campbell G K 2011 \textit{Phys.\ Rev.\ Lett.\ }
\doi{10.1103/PhysRevLett.106.130401}{\textbf{106} 130401}

\bibitem{cassett}
Bruce G D, Mayoh J, Smirne G, Torralbo-Campo L and Cassettari D 2011 \textit{Phys.\ Scr.\ }
\doi{10.1088/0031-8949/2011/T143/014008}{\textbf{T143} 014008}

\bibitem{RubinszteinDunlop:2008:a}
Schnelle S K, van Ooijen E D, Davis M J, Heckenberg N R and Rubinsztein-Dunlop H 2008 \textit{Opt.\ Express} 
\doi{10.1364/OE.16.001405}{\textbf{16}, 1405}

\bibitem{Arnold:2008:b}
Houston N, Riis E and Arnold A S 2008 \textit{J.\ Phys.\ B} 
\doi{10.1088/0953-4075/41/21/211001}{\textbf{41} 211001}

\bibitem{Boshier:2009:a}
Henderson K, Ryu C, MacCormick C and Boshier M G 2009 \textit{New J.\ Phys.\ } 
\doi{10.1088/1367-2630/11/4/043030}{\textbf{11} 043030}

\bibitem{ferris1}
Franke-Arnold S, Leach J, Padgett M J, Lembessis V E, Ellinas D, Wright A J, Girkin J M, \"Ohberg P and 
Arnold A S 2007 \textit{Opt.\ Express} \doi{10.1364/OE.15.008619}{\textbf{15} 8619}

\bibitem{ferris2}
Arnold A S 2012 \textit{Opt.\ Lett.\ }\doi{10.1364/OL.37.002505}{\textbf{37} 2505}




\bibitem{Chapman:2001:a}
Sauer J A, Barrett M D and Chapman M S 2001 \textit{Phys.\ Rev.\ Lett.\ }
\doi{10.1103/PhysRevLett.87.270401}{\textbf{87} 270401}

\bibitem{Prentiss:2004:a}
Wu S, Rooijakkers W, Striehl P and Prentiss M 2004 \textit{Phys.\ Rev.\ A} 
\doi{10.1103/PhysRevA.70.013409}{\textbf{70} 013409}

\bibitem{Riis:2006:a}
Arnold A S, Garvie C S and Riis E 2006 \textit{Phys.\ Rev.\ A} 
\doi{10.1103/PhysRevA.73.041606}{\textbf{73} 041606}

\bibitem{Arnold:2004:a}
Arnold A S 2004 \textit{J.\ Phys.\ B} 
\doi{10.1088/0953-4075/37/2/L03}{\textbf{37} L29}

\bibitem{StamperKurn:2005:a}
Gupta S, Murch K W, Moore K L, Purdy T P and Stamper-Kurn D M 2005 \textit{Phys.\ Rev.\ Lett.\ }
\doi{10.1103/PhysRevLett.95.143201}{\textbf{95} 143201}

\bibitem{Ifan}
West A D, Wade C G, Weatherill K J and Hughes I G 2012 \textit{Appl.\ Phys.\ Lett.\ }
\doi{10.1063/1.4736580}{\textbf{101} 023115}

\bibitem{Fortagh:2002:a}
Kraft S, G\"unther A, Ott H, Wharam D, Zimmermann C and Fort\'agh J 2002 \textit{J.\ Phys.\ B} 
\doi{10.1088/0953-4075/35/21/102}{\textbf{35} L469}

\bibitem{Pritchard:2002:a}
Leanhardt A E, Chikkatur A P, Kielpinski D, Shin Y, Gustavson T L, Ketterle W and Pritchard D E 2002
\textit{Phys.\ Rev.\ Lett.\ } 
\doi{10.1103/PhysRevLett.89.040401}{\textbf{89} 040401}

\bibitem{Pritchard:2003:a}
Leanhardt A E, Shin Y, Chikkatur A P, Kielpinski D, Ketterle W and Pritchard D E 2003 \textit{Phys.\ Rev.\ Lett.\ }
\doi{10.1103/PhysRevLett.90.100404}{\textbf{90} 100404}

\bibitem{Hinds:2003:b}
Jones M P A, Vale C J, Sahagun D, Hall B V and Hinds E A 2003 \textit{Phys.\ Rev.\ Lett.\ }
\doi{10.1103/PhysRevLett.91.080401}{\textbf{91} 080401}

\bibitem{schummcorr}
Schumm T, Est\`{e}ve J, Figl C, Trebbia J-B, Aussibal C, Nguyen H, Mailly D, Bouchoule I, Westbrook C I 
and Aspect A 2005 \textit{Eur.\ Phys.\ J.\ D} \doi{10.1140/epjd/e2005-00016-x}{\textbf{32} 171}

\bibitem{petrich}
Petrich W, Anderson M H, Ensher J R and Cornell E A 1995 \textit{Phys.\ Rev.\ Lett.\ }
\doi{10.1103/PhysRevLett.74.3352}{\textbf{74} 3352}

\bibitem{Garraway:2001:a}
Zobay O and Garraway B M 2001 \textit{Phys.\ Rev.\ Lett.\ }
\doi{10.1103/PhysRevLett.86.1195}{\textbf{86} 1195}

\bibitem{Garraway:2006:a}
Morizot O, Colombe Y, Lorent V, Perrin H and Garraway B M 2006 \textit{Phys.\ Rev.\ A}
\doi{10.1103/PhysRevA.74.023617}{\textbf{74} 023617}

\bibitem{Schmiedmayer:2006:a}
Lesanovsky I, Schumm T, Hofferberth S, Andersson L M, Kr\"uger P and Schmiedmayer J 2006
\textit{Phys.\ Rev.\ A} 
\doi{10.1103/PhysRevA.73.033619}{\textbf{73} 033619}


\bibitem{Spreeuw:2007:a}
Fernholz T, Gerritsma R, Kr\"uger P and Spreeuw R J C 2007 \textit{Phys.\ Rev.\ A} 
\doi{10.1103/PhysRevA.75.063406}{\textbf{75} 063406}

\bibitem{vonKlitzing:2007:a}
Lesanovsky I and von Klitzing W 2007 \textit{Phys.\ Rev.\ Lett.\ }
\doi{10.1103/PhysRevLett.99.083001}{\textbf{99} 083001}

\bibitem{Foot:2008:a}
Heathcote W H, Nugent E, Sheard B T and Foot C J 2008 \textit{New J.\ Phys.\ }
\doi{10.1088/1367-2630/10/4/043012}{\textbf{10} 043012}

\bibitem{Foot:2011}
Sherlock B E, Gildemeister M, Owen E, Nugent E and Foot C J 2011 \textit{Phys.\ Rev.\ A} 
\doi{10.1103/PhysRevA.83.043408}{\textbf{83} 043408}

\bibitem{schmiedRWA}
Hofferberth S, Fischer B, Schumm T, Schmiedmayer J and Lesanovsky I 2007 \textit{Phys. Rev. A} 
\doi{10.1103/PhysRevA.76.013401}{\textbf{76} 013401}


\bibitem{Paul:book:a}
Bergeman T, Erez G and Metcalf H J 1987 \textit{Phys. Rev. A} 
\doi{10.1103/PhysRevA.35.1535}{\textbf{35} 1535}.

\bibitem{Jones}
Jones R C 1941 \textit{J.\ Opt.\ Soc.\ Am}.\ \doi{10.1364/JOSA.31.000488}{\textbf{31} 488} 


\end{thebibliography}
\end{document}